# Using Regression Techniques to Predict Large Data Transfers

Sudharshan Vazhkudai [1]    Jennifer M. Schopf [2]

[1] Department of Computer and Information Science, The University of Mississippi

[2] Mathematics and Computer Science Division, Argonne National Laboratory

{vazhkuda, jms}@mcs.anl.gov

**Abstract**

*The recent proliferation of Data Grids and the increasingly common practice of using resources as distributed data stores provide a convenient environment for communities of researchers to share, replicate, and manage access to copies of large datasets. This has led to the question of which replica can be accessed most efficiently. In such environments, fetching data from one of the several replica locations requires accurate predictions of end-to-end transfer times. The answer to this question can depend on many factors, including physical characteristics of the resources and the load behavior on the CPUs, networks, and storage devices that are part of the end-to-end data path linking possible sources and sinks.*

*Our approach combines end-to-end application throughput observations with network and disk load variations and captures whole-system performance and variations in load patterns. Our predictions characterize the effect of load variations of several shared devices (network and disk) on file transfer times. We develop a suite of univariate and multivariate predictors that can use multiple data sources to improve the accuracy of the predictions as well as address Data Grid variations (availability of data and sporadic nature of transfers). We ran a large set of data transfer experiments using GridFTP and observed performance predictions within 15% error for our testbed sites, which is quite promising for a pragmatic system.*

## 1. Introduction

As the coordinated use of distributed resources, or Grid computing, becomes more commonplace, basic resource usage is changing. Many recent applications use Grid systems as distributed data stores [DataGrid02, GriPhyN02, HSS00, LIGO02, MMR+01, NM02], where pieces of large datasets are replicated over several sites. For example, several high-energy physics experiments have agreed on a tiered Data Grid architecture [HJS+00, Holtman00] in which all data (approximately 20 petabytes by 2006) is located at a single Tier 0 site; various (overlapping) subsets of this data are located at national Tier 1 sites, each with roughly one-tenth the capacity; smaller subsets are cached at smaller Tier 2 regional sites;

and so on. Therefore, any particular dataset is likely to have replicas located at multiple sites [RF01, LSZ+02, LSZ+03].

Different sites may have varying performance characteristics because of diverse storage system architectures, network connectivity features, or load characteristics. Users (or brokers acting on their behalf) may want to be able to determine the site from which particular data sets can be retrieved most efficiently, especially as data sets of interest tend to be large (1–1000 MB). It is this *replica selection* problem that we address in this paper.

Since large file transfers can be costly, there is a significant benefit in selecting the most appropriate replica for a given set of constraints [ACF+02, VTF01]. One way a more intelligent replica selection can be achieved is by having replica locations expose performance information about past data transfers. This information can, in theory, provide a reasonable approximation of the end-to-end throughput for a particular transfer. It can then be used to make predictions about the future behavior between the sites involved. In our work we use GridFTP [AFN+01], part of the Globus Toolkit$^{TM}$ [FK98, Globus02] for moving data, but the approach we present is applicable to other large file transfer tools as well.

In this paper we present two- and three-datastream predictions using regression techniques to predict the performance of GridFTP transfers for large file across the Grid. We start by deriving predictions from past history of GridFTP transfers in isolation. We build a suite of univariate predictors comprising simple mathematical models such as mean- and median-based tools that are easy to implement and achieve acceptable levels of accuracy. We then present a detailed analysis of several variations of our univariate forecasting tools and information on GridFTP logs.

The univariate models do not achieve better prediction accuracy because they fail to account for the sporadic nature of data transfers in Grid environments. Hence, predictions based on simple log data may not contain enough recent information on current system trends. We need to be able to derive forecasts from several combinations of currently available data sources in order to capture information about the current Grid environment.

To address this need, we use both log data and periodic data to expose the behavior of key components in the end-to-end data path. We use the additional datastreams of network and disk behavior to illustrate how additional data can be exploited in predicting the behavior of large transfers. We present an in-depth study of these data sources and our multivariate forecasting tools, including information about data formats, lifetime, time/space constraints, correlation, statistical background on our regression tools, and the advantages and disadvantages of this approach. While in this paper, we have demonstrated univariate and multivariate predictors for the GridFTP tool, nothing in our approach limits us to any single protocol and the predictors can be applied to any wide-area data movement tool.

We then evaluate our prediction approaches using several different metrics. Comparing the normalized percentage errors of our various predictions, we find that the univariate predictions have error rates of at most 25% and that all the univariate predictors performed similarly. With multivariate predictions, we observed that combining GridFTP logs and disk throughput observations provided us with gains of up to 4% when compared with the best of univariate predictors. Combining logs with network throughput data provides further gains up to 6%, and predictions based on all three data sources had up to 9% reduction in error. To study the degree of variance in error rates, we computed confidence levels and observed that the variance is smaller with more accurate predictors for the sites we examined. We further developed a triplet metric comprising the throughput, percentage error rate, and confidence level as a measure of a given site's predictive merit.

## 2. Related Work

The goal of this work is to obtain accurate predictions of file transfer times between a storage system and a client. Achieving this can be challenging because numerous devices are involved in the end-to-end path between the source and the client, and the performance of each (shared) device along the end-to-end path may vary in unpredictable ways.

One approach to predicting this information is to construct performance models for each system component (CPUs at the level of cache hits and disk access, networks at the level of the individual routers, etc.) and then to use these models to determine a schedule for all data transfers [SC00], similar to classical scheduling [Adve93, Cole89, CQ93, Crovella99, ML90, Schopf97, TB86, ZLP96]. In practice, however, it is often unclear how to combine this data to achieve accurate end-to-end measurements. Also, since system components are shared, their behavior can vary in unpredictable ways [SB98]. Further, modeling individual components in a system may not capture the significant effects these components have on each other, thereby leading to inaccuracies [GT99].

Alternatively, observations from past application performance of the entire system can be used to predict end-to-end behavior. The use of whole-system observation has relevant properties for our purposes. These predictions can, in principle, capture both evolution in system configuration and temporal patterns in load. A by-product of capturing entire system evolution is enhanced transparency, in that we can construct such predictions without detailed knowledge of the underlying physical devices. This technique is used by Downey [Downey97] and Smith et al. [SFT98] to predict queue wait times and by numerous tools (Network Weather Service [Wolski98], NetLogger [NetLogger02], Web100 [Web100Project02], iperf [TF01], and Netperf [Jones02]) to predict the network behavior of small file transfers.

Although tools such as the Network Weather Service (NWS) measure and predict network bandwidth, a substantial difference in performance can arise between a small NWS probe (lightweight with 64 KB size) and an actual file transfer using GridFTP (with tuned TCP buffers and parallelism). We show this in Figure 1, which depicts 64 KB NWS measurements, indicating that the bandwidth is about 0.3 MB/sec, and end-to-end GridFTP measurements for files ranging from 1 to 1000 MB in size, indicating a significantly higher transfer rate. In this case, NWS by itself is not sufficient to predict end-to-end GridFTP throughput. In addition, we see a much larger variability in GridFTP measurements, ranging from 1.5 to 10.2 MB/sec (because of different transfer sizes and also load variations in the end-to-end components), so that it is unlikely that a simple data transformation will improve the resulting prediction.

The univariate predictors presented in this work are similar to the basic predictors used by NWS and similar tools to predict the behavior of time series data. Because our data traces are not periodic in nature, however, we also use predictions based on multiple datastreams. This approach is similar to work done by Faerman et al. [FSW+99], which used the NWS and adaptive linear regression models for the Storage Resource Broker [BMR+98] and SARA [SARA02]. Faerman and his colleagues compared transfer times obtained from a raw bandwidth model (`Transfer-Time = ApplicationDataSize/NWS-Probe-Bandwidth`, with 64 KB NWS probes) with predictions from regression models and observed accuracy improvements ranging from 20% to almost 100% for the sites examined. The work presented here goes beyond that work, however, by exploring several filling techniques to mitigate adverse effects of sporadic transfers.

Swany and Wolski have also approached multivariate predictors by constructing cumulative distribution functions of past history and deriving predictions from them as an alternative to regressive models. This approach has been demonstrated for 16 MB HTTP transfers with improved prediction accuracy when compared with their univariate prediction approach [SW02]. Further, they have applied their models to our datasets, comprising various file sizes, and have observed comparable prediction accuracy.

## 3. Data Sources

In this section, we describe our three primary data sources. We use the GridFTP server to perform our data transfers and log its behavior every time a transfer is made, thereby recording the end-to-end transfer behavior. Since these events are very sporadic in nature, however, we also need to capture data about the current environment to have accurate predictions. Hence, we use the Network Weather Service probe data as an estimate of bandwidth for small data transfers and the iostat disk throughput data to measure disk behavior.

### 3.1. GridFTP Logs

GridFTP [AFN+01] is part of the Globus Toolkit™ [FK98, Globus02] and is widely used as a secure, high-performance data transfer protocol [ACF+02, AFN+01, DataGrid02, GriPhyN02]. It extends standard FTP implementations with several features needed in Grid environments, such as security, parallel transfers, partial file transfers, and third party transfers. We instrumented the GT 2.0 wuftp-based GridFTP server to log the source address, file name, file size, number of parallel streams, stripes, TCP buffer size for the transfer, start and end timestamps, nature of the operation (read/write), and logical volume to/from which file was transferred (Table 1) [VSF02].

The GridFTP monitoring code is nonintrusive. The majority of the overhead is in the timing routines, with a smaller percentage spent gathering the information mentioned above and performing a write operation. The entire logging process consumes on average of approximately 25 milliseconds per transfer, which is insignificant compared with the total transfer time.

Although each log entry is well under 512 bytes, transfer logs can grow quickly in size at a busy site. We do not currently implement a log management scheme, but it would be straightforward to use a circular buffer, such as in the NWS. An alternative strategy used by NetLogger is to flush the logs to persistent storage (either disk or network) and restart logging.

### 3.2. NWS

The Network Weather Service [Wolski98] monitors the behavior of various resource components by sending out lightweight probes or querying system files at regular intervals. NWS sensors exist for components such as CPU, disk, and network. We used the network bandwidth sensor with 64 KB probes to estimate the current network throughput. NWS throughput measurements, although not representative of the transfer bandwidth obtainable for large files (10 MB to 1 GB), are representative of the network link characteristics. Further, NWS is intended to be a lightweight, noninvasive monitoring system (only a few milliseconds of overhead) whose measurements can then be extrapolated to specific cases such as ours.

### 3.3. Iostat

Traditionally, in large wide-area transfers, network transport has been considered to weigh heavily on the end-to-end throughput achieved. Current trends in disk storage and networking, however, suggest that disk accesses will factor rather strongly in the future. Network throughput is far outpacing advances in disk speeds. Therefore, as link speeds increase, the network latency significantly drops, and disk accesses are likely to become the bottleneck in large file transfers across the Grid [GS00].

To address this issue, we include disk throughput data in our prediction approach. The iostat tool is part of the sysstat [SYSSTAT02] system-monitoring suite and collects disk I/O throughput data by monitoring the blocks read/written from/to a particular disk. Iostat can be configured to periodically monitor disk transfer rates, block read/write rates, and so forth of all physically connected disks. We use the disk transfer rate that represents the throughput of the disk. This also has an overhead of only a few milliseconds.

### 3.4. Correlation

A key step in analyzing whether a combination of datastreams will result in better predictions is to evaluate how highly correlated they are. The correlation coefficient is a measure of the linear relationship between two variables and can have a value between –1.0 and +1.0 depending on the strength of the relation. A coefficient near zero suggests that the variables may not be linearly related, although they may exhibit nonlinear dependencies [Edwards84, OM88]. The correlation coefficient for two datastreams G and N is computed by using the formula

$$\text{corr} = \frac{\sum NG - (\sum N \sum G / \text{size})}{\sqrt{(\sum G^2 - (\sum G)^2 / \text{size})} \ \sqrt{(\sum N^2 - (\sum N)^2 / \text{size})}},$$

where *size* is the number of values in the data stream.

We compute the rank-order correlation for each of our datasets. Rank correlation provides a distribution-free, nonparametric alternative to determine whether the observed correlation is significant [Edwards84]. Rank correlation converts data to ranks by assigning a specific rank to each value in the datastream, as determined by the position of the value when the datastream is sorted. Table 2 shows a tabulated listing of the 95% confidence interval for the correlation coefficients for the three datasets we collected between our transfer points. The confidence interval denotes that the correlation for 95% of the sample falls within a certain upper and lower limit. We can see a moderate correlation between GridFTP, NWS, and disk throughput datastreams.

## 4. Predictors

We evaluated a wide set of prediction techniques for wide-area data transfers. This section presents the univariate predictions and the multivariate prediction techniques we used in our experiments.

### 4.1. Univariate Predictors

In this section we describe some of the predictors we developed, categorize possible approaches by basic mathematical techniques, and detail the advantages and disadvantages of each technique.

#### 4.1.1. Mathematical Functions

Mathematical functions for predictions are generally grouped into mean-based, median-based, and autoregressive techniques. We use several variations of each of these models in our experiments.

Mean-based, or averaging, techniques are a standard class of predictors that use arithmetic averaging (as an estimate of the mean value) over some portion of the measurement history to estimate future behavior. The general formula for these techniques is the sum of the previous *n* values over the number of measurements. Mean-based predictors are easy to implement and impose minimally on system resources.

A second class of standard predictors is based on evaluating the median of a set of values. Given an ordered list of *t* values, if *t* is odd, the median is the $(t+1)/2$ value; if *t* is even, the median is half of the $t/2$ value added with the $(t+1)/2$ value. Median-based predictors are particularly useful if the measurements contain randomly occurring asymmetric outliers that are rejected. However, they lack some of the smoothing that occurs with a mean-based method, possibly resulting in forecasts with a considerable amount of jitter [HP91].

The third class of common predictors is autoregressive models [GP94, HP91, Wolski98]. We use an autoregressive integrated moving average (ARIMA) model technique that is constructed using the equation

$$G' = a + bG_{t-1},$$

where G' is the GridFTP prediction for time, t, $G_{t-1}$ is the previous data occurrence, and *a* and *b* are the regression coefficients that are computed based on past occurrences of G using the method of least squares. This approach is most appropriate when there are at least fifty measurements and the data is measured with equally spaced time intervals. Our data does not meet these constraints, but we include this technique to do a full comparison. The main advantage of using an ARIMA model is that it gives a weighted average of the past values of the series, thereby possibly giving a more accurate prediction. However, in addition to requiring a larger data set than the other techniques to achieve a statistically significant result, the model can have a much greater computational cost.

### 4.1.2. Context-Insensitive Variants

When evaluating a data set, the values can be filtered in several ways to include only data that is relevant to the current prediction environment. We evaluate two general ways of altering our base formulas to do filtering: *context-insensitive variants* that include data independent of the meaning of the data, primarily temporal filtering, and *context-sensitive variants*, in which data is culled based on the context of the values.

More recent values are often better predictors of future behavior than an entire dataset, no matter which mathematical technique is used to calculate a prediction. Hence, many different variants exist in selecting a set of recent measurements to use in a prediction calculation, creating several different context-sensitive variants on our original prediction models.

The fixed-length, or sliding window, average is calculated by using only a set number of previous measurements to calculate the average. The number of measurements can be chosen statically or dynamically depending on the system. We use only static selection techniques in this work. Options for dynamically selecting window size are discussed in [Wolski98]. The degenerative case of this strategy involves using only the last measurement to predict the future behavior. Work by Downey and Harchol-Balter [HD96] shows that this is a useful predictor for CPU resources, for example.

Instead of selecting the number of recent measurements to use in a prediction, we also consider using only a set of measurements from a previous window of time. Unlike other systems where measurements are taken at regular intervals [DO00, Wolski98], our measurements can be spaced irregularly. Using temporal windows for irregular samples can reflect trends more accurately than selecting a specific number of previous measurements because they capture recent fluctuations, thereby helping to ensure that recent (and, one hopes, more predictive) data is used. Much as the number of measurements included in a prediction can be selected dynamically, the window of time used can be decided dynamically.

As shown in Table 3, we use fixed-length sets of the last 1 (last value), 5, 15, and 25 measurements. We use temporal-window sets of data of the last 5 hours, 15 hours, 25 hours, 5 days, and 10 days. We consider both mean-based and median-based predictors over previous *n* measurements; mean-based predictors over the previous 5, 15, and 25 hours; and autoregression (AR) over the previous 5 and 10 days, since this function requires a much larger dataset to produce accurate predictions than our other techniques.

### 4.1.3. Context-Sensitive Variants

Filtering a data set to eliminate unrelated values often results in a more accurate prediction. For example, a prediction of salary is more accurate when factors such as previous training, education, and years at the position are used to limit the dataset of interest. By doing this we generate several context-sensitive variants of our original prediction models.

With the GridFTP monitoring data, initial results showed that file transfer rates had a strong correlation with file size. Studies of Internet traffic have also revealed that small files achieve low bandwidths whereas larger files tend to have high bandwidths [BMK96, CSA98, GM01]. This difference is thought to be primarily due to the startup overhead associated with the TCP start mechanism that probes the bandwidth at connection startup. Recent work has focused on class-based isolation of TCP flows [YM01] and on startup optimizations [ZQK00, ZQK99] to mitigate this problem. As a proof of concept, we found 5%–10% improvement on average when using file-size classification instead of the entire history file to calculate a prediction. This is discussed in Section 5.

For our GridFTP transfer data we ran a series of tests between our testbed sites to categorize the data sizes into a small number of classes. We categorized our data into four sets: 0–50 MB, 50–250 MB, 250–750 MB, and more than 750 MB based on the achievable bandwidth. We note that these classes apply to the set of hosts for our testbed only; further work is needed to generalize this notion.

### 4.2. Multivariate Predictors

The obvious downside of univariate predictors has nothing to do with the predictors themselves but more so with the nature of data transfers on the Grid. Because of the sporadic nature of transfers, predictors based on log data alone may fail to factor in current system trends and fluctuations. To mitigate the adverse effects of this problem, we introduce other periodic datastreams to expose the behavior of components in the end-to-end data path and to reveal the current environment on the Grid.

We developed a set of multivariate predictors using regression models to predict from a combination of several data sources – GridFTP log data and network load data, GridFTP log data and disk load data, or a combination of all three. The datastreams require some preprocessing before the regression techniques can be applied to them. This includes time matching the data streams and filling-in techniques.

### 4.2.1 Matching

Our three data sources (GridFTP, disk I/O, and NWS network data) are collected exclusive of each other and rarely have the same timestamps. To use regressive models on the data streams, however, we need to have a one-to-one mapping for the values in each stream. Hence, we are required to match values from the three sets such that for each GridFTP value, we find disk I/O and network observations that were made around the same time.

For each GridFTP data point ($T_G$, G), we match a corresponding disk load ($T_D$, D) and NWS data point ($T_N$, N) such that $T_N$ and $T_D$ are the closest to $T_G$. By doing this, the triplet ($N_i$, $D_j$, $G_k$) represents an observed end-to-end GridFTP throughput ($G_k$) resulting from a data transfer that occurred with the disk load ($D_j$) and network probe value ($N_i$).

At the end of the matching process, the three datastreams have been combined into the sequence that looks like
$$(N_i, D_j, G_k) (N_{i+1}, D_{j+1}, \_) \ldots (N_{i+m}, D_{j+m}, G_{k+1}),$$
where $G_k$, and $G_{k+1}$ are two successive GridFTP file transfers, $N_i$ and $N_{i+m}$ are NWS measurements, and $D_j$ and $D_{j+m}$ are disk load values that occurred in the same timeframe as the two GridFTP transfers. The sequence also consists of a number of disk load and NWS measurements between the two transfers for which there are no equivalent GridFTP values, such as ($N_{i+1}$, $D_{j+1}$, $\_$). Note that these interspersed network and disk load values are time-aligned. Also note that we have described the matching process with reference to all three data sources. In the case where a prediction uses a different number of datastreams, similar matching techniques can be employed.

### 4.2.2. Filling-in Techniques

After matching the datastreams, we need to address the tuples that do not have values for the GridFTP data – that is, the NWS data or disk I/O data collected in between the sporadic GridFTP transfers. Regression models expect a one-to-one mapping between the data values, so we can either discard the network and I/O data for which there are no equivalent GridFTP data (our NoFill technique, Figure 2) or fill in synthetic transfer values using either an average over the past day's data (Avg), or the last value (LV). Once filled in, the sequence is as follows:
$$(N_i, D_j, G_k) (N_{i+1}, D_{j+1}, G_{Fill}) \ldots (N_{i+m}, D_{j+m}, G_{k+1})$$

where $G_{Fill}$ is the synthetic GridFTP value. Data, once matched and filled in, is fed to regression models (Figure 3).

### 4.2.3. Linear Regression

Linear regression attempts to build linear models between dependent and independent variables. The following equation builds linear models between several independent variables $N_1, N_2, ..., N_k$ and dependent variable G as follows:

$$G' = a + b_1 N_1 + b_2 N_2 + ... + b_k N_k,$$

where `G'` is the prediction of the observed value of `G` for the corresponding values of $N_1, N_2, ..., N_k$. The coefficients `a`, $b_1$, $b_2$, and $b_k$ are calculated by using the method of least squares [Edwards84]. For our case, we built linear models between NWS (`N`), disk (`D`), and GridFTP (`G`) data as explained above, with `N` and `D` as independent variables.

### 4.2.4. Polynomial Regression Models

To improve prediction accuracy, we also developed a set of nonlinear models adding polynomial terms to the linear equation. For instance, a quadratic model is as follows:

$$G' = a + b_1 N + b_2 N^2.$$

Cubic and quartic models have additional terms $b_3 N^3$ and $b_4 N^4$, respectively. Similar to the linear model, the coefficients in quadratic, cubic, and quartic models $b_2$, $b_3$, and $b_4$ are computed by using the method of least squares. Adding polynomial terms to the regression model can reach a saturation point (no significant improvement in prediction accuracy observed), suggesting that a particular model sufficiently captures the relationship between the two variables [OM88, Pankratz91]. Figure 4 shows a bar graph that compares error, complexity of algorithm, and components included for the site pair, Lawrence Berkeley and Argonne National Laboratories.

## 5. Measurements and Evaluation

We evaluated the performance of our regression techniques on datasets collected over three distinct two-week durations: August 2001, December 2001, and January 2002. In the following subsections we describe the experimental setup, prediction error calculations, and the results obtained from these datasets.

### 5.1. Experimental Setup

The experiments we ran consisted of controlled GridFTP transfers, NWS network sensor measurements, and disk throughput monitoring between four sites in our testbed (Figure 5): Argonne National Laboratory (ANL), the University of Southern California Information Sciences Institute (ISI), Lawrence Berkeley National Laboratory (LBL) and the University of Florida at Gainesville (UFL).

GridFTP experiments included transfers comprising several file sizes ranging from 10 MB to 1 GB, performed at random time intervals within 12-hour periods. We calculated buffer sizes by using the formula

```
         RTT * "bottleneck bandwidth in the link"
```
with roundtrip times (RTT) values obtained from ping and with bottleneck bandwidth obtained by using iperf [TF01]. Figure 5 shows the roundtrip times and bottleneck bandwidth for our site pairs. Our GridFTP experiments were performed with tuned TCP buffer settings (1 MB based on the bandwidth delay product) and eight parallel streams to achieve enhanced throughput. Logs of these transfers were maintained at the respective sites and can be found at [Traces02].

Configuring NWS among a set of resources involved setting up a nameserver and memory to which sensors at various sites registered and logged measurements [Wolski98]. In our experiments, we used ANL as a registration and memory resource. NWS network monitoring sensors between these sites were set up to measure bandwidth every five minutes with 64 KB probes.

Disk I/O throughput data was collected by using the iostat tool logging transfer rates every five minutes. Logs were maintained at the respective servers.

For each data set and predictor, we used a 15-value training set; that is, we assumed that at the start of a predictive technique there were at least 15 GridFTP values in the log file (approximately two days worth of data).

### 5.2. Metrics

We calculate the prediction accuracy using the normalized percentage error calculation:

```
            Σ | MeasuredBW – PredictedBW |
 % Error = ─────────────────────────────── * 100,
                  (size * MeanBW)
```

where *size* is the total number of predictions and the Mean is the average measured GridFTP throughput. In this subsection we show results based on the August 2001 dataset. Complete results for all three datasets can be found in the appendix and at [Traces02].

In addition to evaluating the error of our predictions, we evaluate information about the variance in the error. Depending on the use case, a user may be more interested in selecting a site that has reasonable performance bandwidth estimates with a relatively low prediction error than in selecting a resource with higher performance estimates and a possibly much higher error in prediction. In such cases, it can be useful if the forecasting error can be stated with some confidence and with a maximum/minimum variation range. These limits can also, in theory, be used as catalysts for corrective measures in case of performance degradation.

In our case, we can also use these limits to verify the inherent cost of accuracy of the predictors. By comparing the confidence intervals of these prediction error rates, we can determine whether the accuracy achieved is at the cost of greater variability, in which case there is little gain in increasing the component complexity of our prediction approach.

Thus, for any predictor (for any site pair and a given dataset), the information denoted by the following triplet can be used as a metric to gauge its accuracy:

```
Accuracy-Metric = [PredictedThroughput, AvgPast % Error-Rate,
                   ConfidenceLimit],
```

where *PredictedThroughput* is the predicted GridFTP value (higher the better), with a certain percentage prediction error (the lower the better) and a percentage confidence interval for the error (the smaller the better).

### 5.3. Univariate Predictor Performance

Figure 6 shows bar charts of percentage error for our various univariate predictors at the various site pairs. The major result from these predictions is that even simple techniques have a worst-case prediction of about 25%, quite respectable for pragmatic prediction systems.

Figure 7 shows the result of sorting the data by file size, since GridFTP throughput varied with transfer file sizes. We grouped several file sizes into categories: 0–50 MB as 10M, 50–250 MB as 100M, 250–750 MB as 500M, and more than 750 MB as 1G, based on the achievable bandwidth. We observe almost up to 10% increase in accuracy with context sensitive filtering.

Figure 8 shows the relative performance of the predictors to determine which predictor performed better by computing the best and worst predictor for each data transfer. On average, predictors that had a high best percentage also had a high worst percentage.

In general, for our univariate predictors, we did not see a noticeable advantage of limiting either average or median techniques using a sliding window or time frames. The ARIMA models did not see improved performance for our data, although they are significantly more expensive compared to simple means and medians. This is likely due to the irregular nature of our data. Average and median based predictors (and their temporal variants) for a GridFTP dataset size of 50 values was computed under a millisecond, while autoregression on the same set consumed a few milliseconds.

### 5.4. Multivariate Predictor Performance

Table 4 shows the performance gains of using a regression prediction with GridFTP and NWS network data (G+N) over using the GridFTP log data univariate predictor in isolation (first two shaded columns in the table). We use the moving average (AVG25) as a representative of univariate predictor performance. For our datasets, we observed a 4% to 6% improvement in prediction accuracy when the regression techniques with LV or AVG filling were used. Regression with NoFill (throwing away the unmatched GridFTP data) shows no significant improvement when compared with univariate predictors.

Table 4 also shows that including disk I/O component load variations in the regression model provides us with gains of 2% to 4% (G+D Avg) when compared with AVG25 (first and third shaded columns in the table). Different filling techniques (G+D Avg and G+D LV) perform similarly, and again NoFill shows no improvement, or even a decrease in accuracy, when compared with univariate predictors.

Comparing the second and third block of data in Table 4 shows that all variations of predictors using NWS data (G+N) perform better than predictors using disk I/O data (G+D) in general. This observation agrees with our initial measurements that only 15% to 30% of the total transfer time is spent in I/O, while the majority of the transfer time (in our experiments) is spent performing network transport.

When we include both disk I/O and NWS network data in the regression model (G+N+D) along with GridFTP transfer logs, we see prediction error drop of 8% to 17% and up to 3% improvement when compared with G+N (second and fourth shaded columns in Table 4) and a 6% improvement over G+D (third and fourth shaded columns in Table 4). Overall, we see up to 9% improvement when we compare G+N+D with the original univariate prediction based on AVG25.

Figure 9a compares the average prediction error for Moving Avg, G+D Avg, G+N Avg, and G+N+D Avg for all of our site pairs (represents the shaded columns in Table 4) and also presents 95% confidence limits for our prediction error rates. The prediction accuracy trend is as follows:
```
Moving Avg < (G+D Avg) < (G+N Avg) < (G+N+D Avg)
```

Figure 9b shows that the confidence interval (the variance in the error) does in fact reduce with more accurate predictors, but the reduction is not significant for our datasets.

Figure 10 depicts the performance of predictors G+D Avg, G+N Avg and G+N+D Avg. The predictors closely track the measured GridFTP values. Predictions were obtained by using regression equations that were computed for each observed network or disk throughput value.

For our datasets, we observed no noticeable improvement in prediction accuracy by using polynomial models for our site pairs. We studied the effects of polynomial regression on all our multivariate tools (G+D, G+N and G+N+D). Figure 11 shows the performance of linear, quadratic, cubic, and quartic regression models for various site pairs for the G+D Avg predictor. All our models performed similarly. On average, regression-based predictors with filling took approximately 10 msec for a dataset size of 50 GridFTP, 1500 NWS values and 1500 iostat values, so are more compute-intensive than univariate models, although still extremely lightweight when compared to the time to transfer the files.

## 6.     Conclusions

In this paper we describe the need for predicting the performance of GridFTP data transfers in the context of replica selection in Data Grids. We show how bulk data transfer predictions can be derived and how its accuracy can be improved by including information on current system/network trends. Further, we argue how data transfer predictions can be constructed using several combinations of datasets. We detail the development of a suite of univariate and multivariate predictors that satisfy the specific constraints of Data Grid environments. We examine a series of simple univariate predictors that are lightweight and use means, medians, and autoregressive techniques. We observed that sliding-window variants tend to capture trends in throughput better than simple means and medians. We also use more complex regression analysis for multivariate predictors. To mitigate the adverse effects of sporadic transfers, multivariate predictors use several filling-in techniques such as last value (LV) and average (AVG). We observe that multivariate predictors with filling offer considerable benefits (up to 9%) when compared with univariate predictors and all our predictors performed better when forecasts were based on clusters of file classifications.

In the future, we are considering integration of our prediction tools into the Data Grid middleware so users and brokers can query them for estimates. The prediction service could, for instance, choose a predictor on-the-fly as the case demands and provide recommendations for possible alternatives. This predictor could easily be written in such a way that it would not be tied to a particular data movement tool.

**Appendix**

Tables 5 and 6 show the performance gains of using a regression prediction with GridFTP and NWS network data (G+N) over using the GridFTP log data univariate predictor in isolation for the December 2001 and January 2002 datasets. Behaviors of both univariate and multivariate predictors are similar to those exhibited in the August 2001 dataset (Table 4). In general, we observe performance improvements in using regression-based filling predictors and prediction error reduces with the addition of disk and network data.

**Acknowledgments**

We thank all the system administrators of our testbed sites for their valuable assistance. This work was supported in part by the Mathematical, Information, and Computational Sciences Division subprogram of the Office of Advanced Scientific Computing Research, Office of Science, U.S. Department of Energy, under contract W-31-109-Eng-38.

**Author Biography**

**Sudharshan S. Vazhkudai** is a doctoral candidate in the Computer Science Department at the University of Mississippi and received a Givens fellowship and a dissertation fellowship to work at Argonne National Laboratory during the summer of 2000 and the academic years 2001 and 2002. He received his bachelors in computer science from Karnatak University, India in 1996, and a masters degree in computer science from The University of Mississippi in 1998. His master's thesis addressed the construction of performance-oriented distributed OS. His current research interest is in distributed resource management.

**Jennifer M. Schopf** received the BA degree in Computer Science and Mathematics from Vassar College in 1992. She received the MS and PhD degrees from the University of California, San Diego, in 1994 and 1998 respectively, in Computer Science and Engineering. While at UCSD she was a member of the AppLeS project. Currently, she is an assistant computer scientist at the Distributed Systems Lab, part of the Mathematics and Computer Science Division at Argonne National Lab, where she is a member of the Globus Project. She also holds a fellow position with the Computational Institute at the University of Chicago and Argonne National Lab and a visiting faculty position at the University of Chicago, Computer Science Department. Her research is in the area of monitoring, performance prediction, and resource scheduling and selection. She is on the steering group of the Global Grid Forum as the area co-director for the Scheduling and Resource Management Area. She is also a co-editor for the upcoming book "Resource Management for Grid Computing", Kluwer, Fall 2003.

**Table 1:** Sample set from a log of file transfers between Argonne and Lawrence Berkeley National Laboratories. The bandwidth values logged are sustained measures through the transfer. The end-to-end GridFTP bandwidth is obtained by the formula BW = file size / transfer time.

| Source IP | File Name | File Size (Bytes) | Volume | StartTime (Timestamp) | EndTime (Timestamp) | TotalTime (Seconds) | Bandwidth (KB/Sec) | Read/Write | Streams | TCP-Buffer |
|---|---|---|---|---|---|---|---|---|---|---|
| 140.221.65.69 | /home/ftp/vazhkuda/10 MB | 10240000 | /home/ftp | 998988165 | 998988169 | 4 | 2560 | Read | 8 | 1000000 |
| 140.221.65.69 | /home/ftp/vazhkuda/25 MB | 25600000 | /home/ftp | 998988172 | 998988176 | 4 | 6400 | Read | 8 | 1000000 |
| 140.221.65.69 | /home/ftp/vazhkuda/50 MB | 51200000 | /home/ftp | 998988181 | 998988190 | 9 | 5688 | Read | 8 | 1000000 |
| 140.221.65.69 | /home/ftp/vazhkuda/100 MB | 102400000 | /home/ftp | 998988199 | 998988221 | 22 | 4654 | Read | 8 | 1000000 |
| 140.221.65.69 | /home/ftp/vazhkuda/250 MB | 256000000 | /home/ftp | 998988224 | 998988256 | 33 | 8000 | Read | 8 | 1000000 |
| 140.221.65.69 | /home/ftp/vazhkuda/500 MB | 512000000 | /home/ftp | 998988258 | 998988335 | 67 | 7641 | Read | 8 | 1000000 |
| 140.221.65.69 | /home/ftp/vazhkuda/750 MB | 768000000 | /home/ftp | 998988338 | 998988425 | 97 | 7917 | Read | 8 | 1000000 |
| 140.221.65.69 | /home/ftp/vazhkuda/1 GB | 1024000000 | /home/ftp | 998988428 | 998988554 | 126 | 8126 | Read | 8 | 1000000 |

**Table 2:** 95% Confidence for the upper and lower limits of the rank-order correlation coefficient for the GridFTP, NWS, and disk I/O datasets between four sites in our testbed. Denotes coefficients for our three datasets.

|  | GridFTP and NWS | | | | | | GridFTP and Disk I/O | | | | | |
|---|---|---|---|---|---|---|---|---|---|---|---|---|
|  | Aug'01 | | Dec'01 | | Jan'02 | | Aug'01 | | Dec'01 | | Jan'02 | |
|  | *Upper* | *Lower* | *Upper* | *Lower* | *Upper* | *Lower* | *Upper* | *Lower* | *Upper* | *Lower* | *Upper* | *Lower* |
| LBL-ANL | 0.8 | 0.5 | 0.5 | 0.3 | 0.6 | 0.2 | 0.6 | 0.1 | 0.5 | 0.2 | 0.5 | 0.1 |
| LBL-UFL | 0.7 | 0.5 | 0.7 | 0.4 | 0.6 | 0.1 | 0.5 | 0.2 | 0.5 | 0.3 | 0.5 | 0.3 |
| ISI-ANL | 0.8 | 0.5 | 0.6 | 0.4 | 0.7 | 0.3 | 0.5 | 0.2 | 0.6 | 0.4 | 0.6 | 0.3 |
| ISI-UFL | 0.9 | 0.4 | 0.6 | 0.2 | 0.5 | 0.1 | 0.5 | 0.1 | 0.6 | 0.3 | 0.5 | 0.2 |
| ANL-UFL | 0.5 | 0.2 | 0.6 | 0.2 | 0.6 | 0.1 | 0.5 | 0.2 | 0.4 | 0.1 | 0.4 | 0.2 |

**Table 3:** Context-insensitive predictors used

|  | **Average based** | **Median based** | **Autoregression** |
|---|---|---|---|
| **All data** | AVG | MED | AR |
| **Last 1 Value** | LV | | |
| **Last 5 Values** | AVG5 | MED5 | |
| **Last 15 Values** | AVG15 | MED15 | |
| **Last 25 Values** | AVG25 | MED25 | |
| **Last 5 Hours** | AVG5hr | | |
| **Last 15 Hours** | AVG15hr | | |
| **Last 25 Hours** | AVG25hr | | |
| **Last 5 Days** | | | AR5d |
| **Last 10 Days** | | | AR10d |

**Table 4:** Normalized percent prediction error rates for the testbed site pairs for the August 2001 dataset. The figure denotes four categories: (1) prediction based on GridFTP data in isolation (AVG25), (2) regression between GridFTP and NWS network data with the three filling in techniques (G+N), (3) regression between GridFTP and disk I/O data with the three filling in techniques (G+D), and (4) regression based on all three data sources (G+N+D). Shaded portions indicate a "best of class" comparison between the approaches. All percentage values are averages based on different file categories.

|  | Only GidFTP Logs [VSF02] | Linear Regression between GridFTP Logs and Network Load [VS02] | | | Linear Regression between GridFTP Logs and Disk Load | | | Linear Regression using all Three Data Sources | | |
|---|---|---|---|---|---|---|---|---|---|---|
|  | AVG25 | G+N NoFill | G+N LV | G+N Avg | G+D NoFill | G+D LV | G+D Avg | G+N+D NoFill | G+N+D LV | G+N+D Avg |
| LBL-ANL | 24.4% | 22.4% | 20.6% | 20% | 25.2% | 21.7% | 21.4% | 22.3% | 17.7% | 17.5% |
| LBL-UFL | 15% | 18.8% | 11.1% | 11% | 20.1% | 11.6% | 11.9% | 11.1% | 8.7% | 8% |
| ISI-ANL | 15% | 12% | 9.5% | 9% | 13.1% | 13% | 11.4% | 11% | 8.9% | 8.3% |
| ISI-UFL | 21% | 21.9% | 16% | 14.5% | 22.7% | 19.7% | 18.8% | 14.7% | 13% | 12% |
| ANL-UFL | 20% | 21% | 20% | 16% | 21.8% | 19.9% | 19.3% | 15.3% | 16.7% | 15.5% |

**Table 5:** Normalized percent prediction error rates for the various site pairs for December 2001 dataset. Figure denotes four categories: (1) prediction based on GridFTP data in isolation (Moving Avg), (2) regression between GridFTP and NWS network data with the three filling in techniques (G+N), (3) regression between GridFTP and disk I/O data with the three filling in techniques (G+D), and (4) regression based on all three data sources (G+N+D). Shaded portions indicate a comparison between our approaches. All percentage values are averages based on different file categories.

|  | Only GidFTP Logs | Linear Regression between GridFTP Logs and Network Load | | | Linear Regression between GridFTP Logs and Disk Load | | | Linear Regression using all Three Data Sources | | |
|---|---|---|---|---|---|---|---|---|---|---|
|  | Moving Avg | G+N NoFill | G+N LV | G+N Avg | G+D NoFill | G+D LV | G+D Avg | G+N+D NoFill | G+N+D LV | G+N+D Avg |
| LBL-ANL | 20% | 23% | 17.6% | 17% | 24% | 19.5% | 19% | 20% | 15.2% | 15.4% |
| LBL-UFL | 16% | 17% | 14.7% | 13% | 16% | 14% | 14.8% | 14.5% | 12.2% | 12% |
| ISI-ANL | 13% | 12% | 10.6% | 9.8% | 12.2% | 11.3% | 11% | 11.3% | 9% | 8.7% |
| ISI-UFL | 17% | 19.3% | 13.2% | 12% | 18% | 15% | 12% | 15% | 10% | 10.8% |
| ANL-UFL | 18% | 18.7% | 14.8% | 14% | 17.8% | 17% | 16.7% | 15.6% | 14% | 13.3% |

**Table 6:** Normalized percent prediction error rates for the various site pairs for January 2002 dataset. Figure denotes four categories: (1) prediction based on GridFTP data in isolation (Moving Avg), (2) regression between GridFTP and NWS network data with the three filling in techniques (G+N), (3) regression between GridFTP and disk I/O data with the three filling in techniques (G+D), and (4) regression based on all three data sources (G+N+D). Shaded portions indicate a comparison between our approaches. All percentage values are averages based on different file categories.

|  | Only GidFTP Logs | Linear Regression between GridFTP logs and network load | | | Linear Regression between GridFTP logs and disk load | | | Linear Regression using all three data sources | | |
|---|---|---|---|---|---|---|---|---|---|---|
|  | Moving Avg | G+N NoFill | G+N LV | G+N Avg | G+D NoFill | G+D LV | G+D Avg | G+N+D NoFill | G+N+D LV | G+N+D Avg |
| LBL-ANL | 26% | 26.8% | 25.5% | 23% | 27% | 25% | 24.8% | 23% | 21.1% | 20.3% |
| LBL-UFL | 21% | 21 | 17.2% | 17% | 23.4% | 21.3% | 20.1% | 17.5% | 14% | 13.3% |
| ISI-ANL | 20% | 19% | 16% | 15.4% | 22.5% | 19% | 19.2% | 19% | 13.6% | 11.8% |
| ISI-UFL | 18% | 18.8% | 13% | 12% | 18.7% | 16.8% | 16.6% | 15% | 10.5% | 11% |
| ANL-UFL | 17% | 19.2% | 12% | 12.2% | 19.2% | 15.7% | 15.9% | 14.1% | 12% | 12.2% |

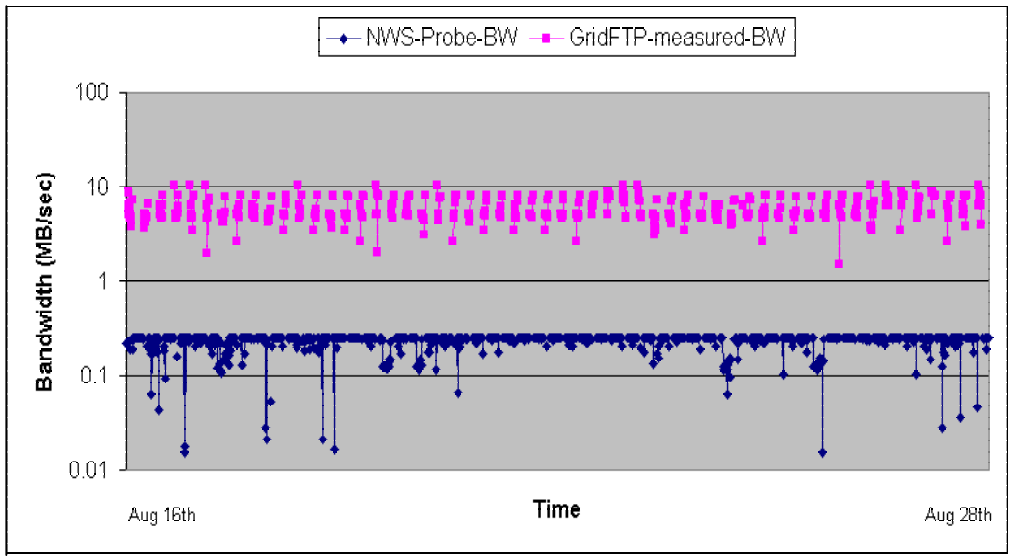

**(a) LBL-ANL**

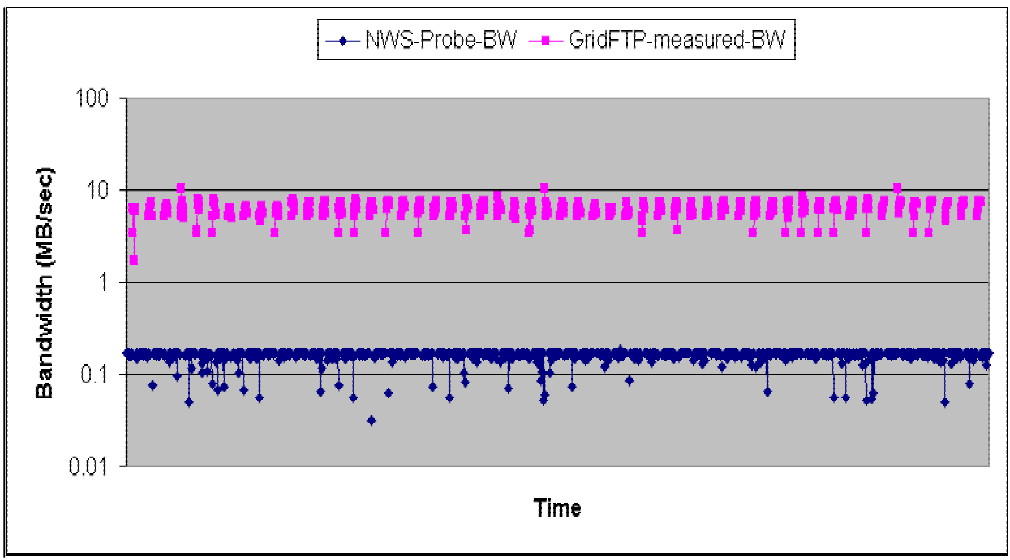

**(b) ISI-ANL**

**Figure 1:** (a) LBL-ANL GridFTP (approximately 400 transfers at irregular intervals) end-to-end bandwidth and NWS (approximately 1,500 probes every five minutes) probe bandwidth for the two-week August'01 dataset. (b) GridFTP transfers and NWS probes between ISI-ANL

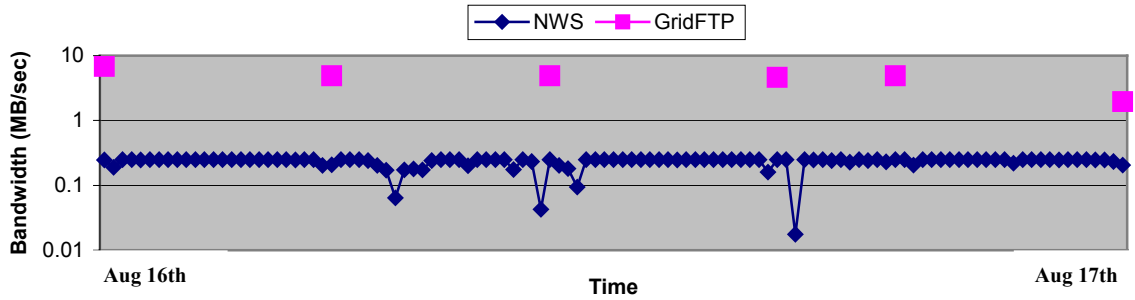

**(a) Measured GridFTP and NWS**

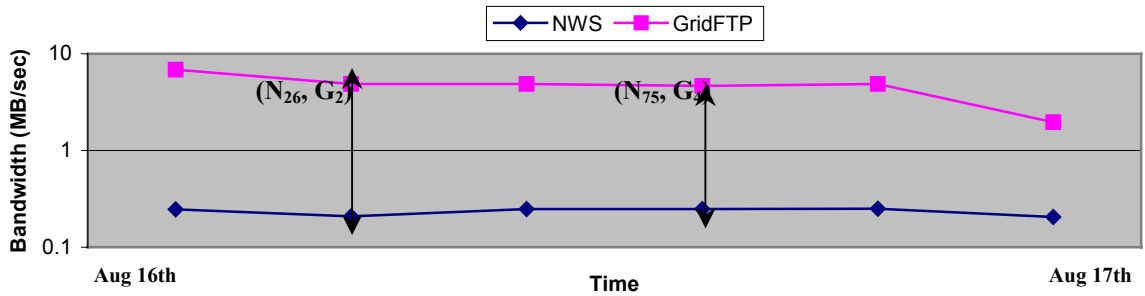

**(b) NoFill**

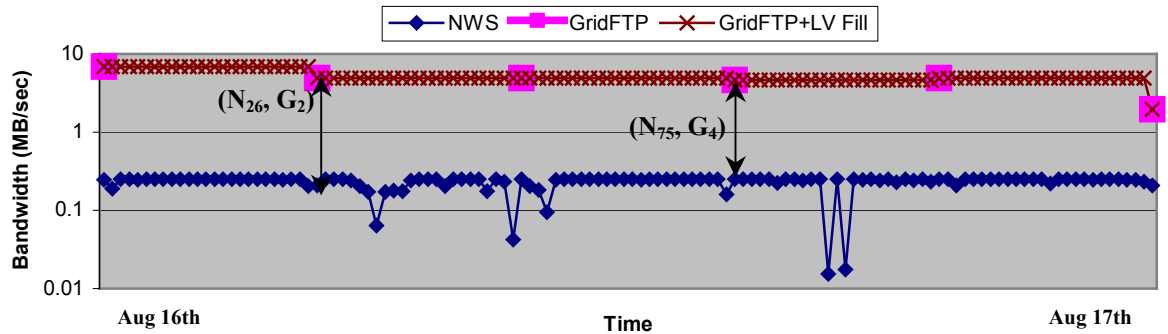

**(c) Last Value Filling (LV)**

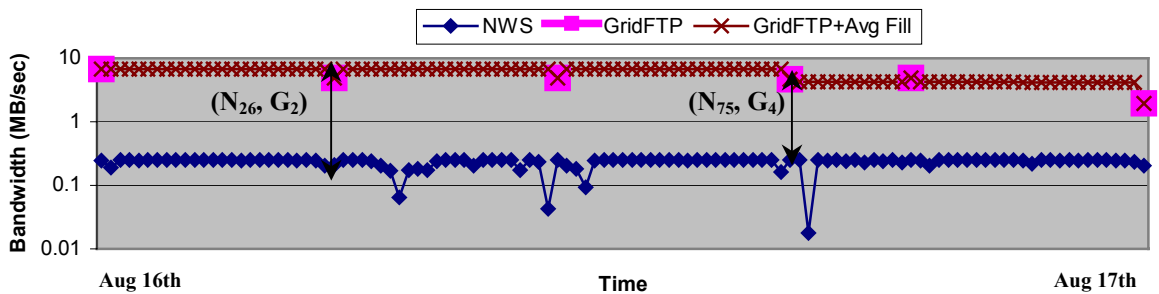

**(d) Average Filling (Avg)**

**Figure 2:** (a) Six measured successive GridFTP transfers and NWS observations during those transfers between LBL and ANL (August 2001). (b) Discarding NWS values to match GridFTP transfers. Here $(N_{26}, G_2)$ denotes that the 26$^{th}$ NWS measurement and the 2$^{nd}$ GridFTP transfer occur in the same timeframe. (c) Filling-in the last GridFTP value to match NWS values between six successive file transfers. (d) Filling-in average of previous GridFTP transfers to match NWS values.

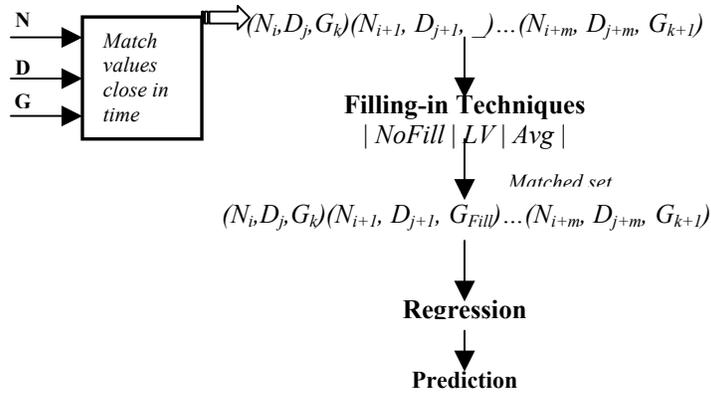

**Figure 3:** Sequence of events for deriving predictions from GridFTP (G), disk load (D), and NWS (N) datastreams.

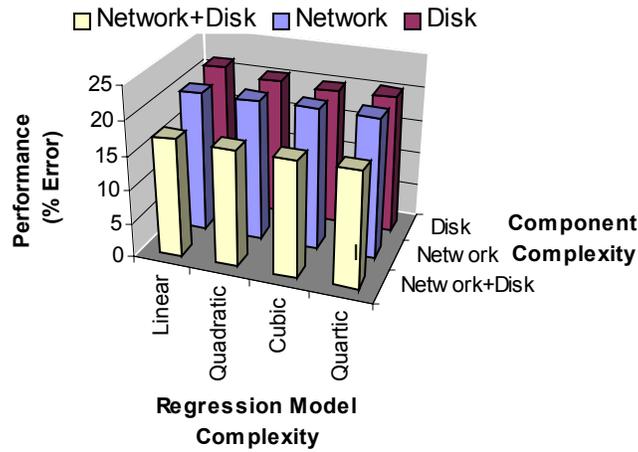

**Figure 4:** Visualization comparing error, complexity of algorithm, and components included for the site pair LBL and ANL.

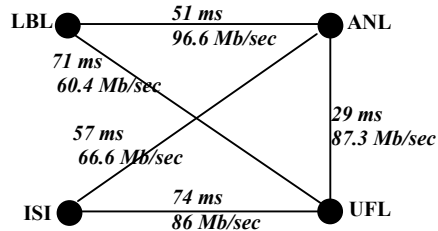

**Figure 5:** Depiction of network settings for our testbed sites connected through OC-48 network links. For each site pair, roundtrip times and bottleneck bandwidths for the link between them is shown.

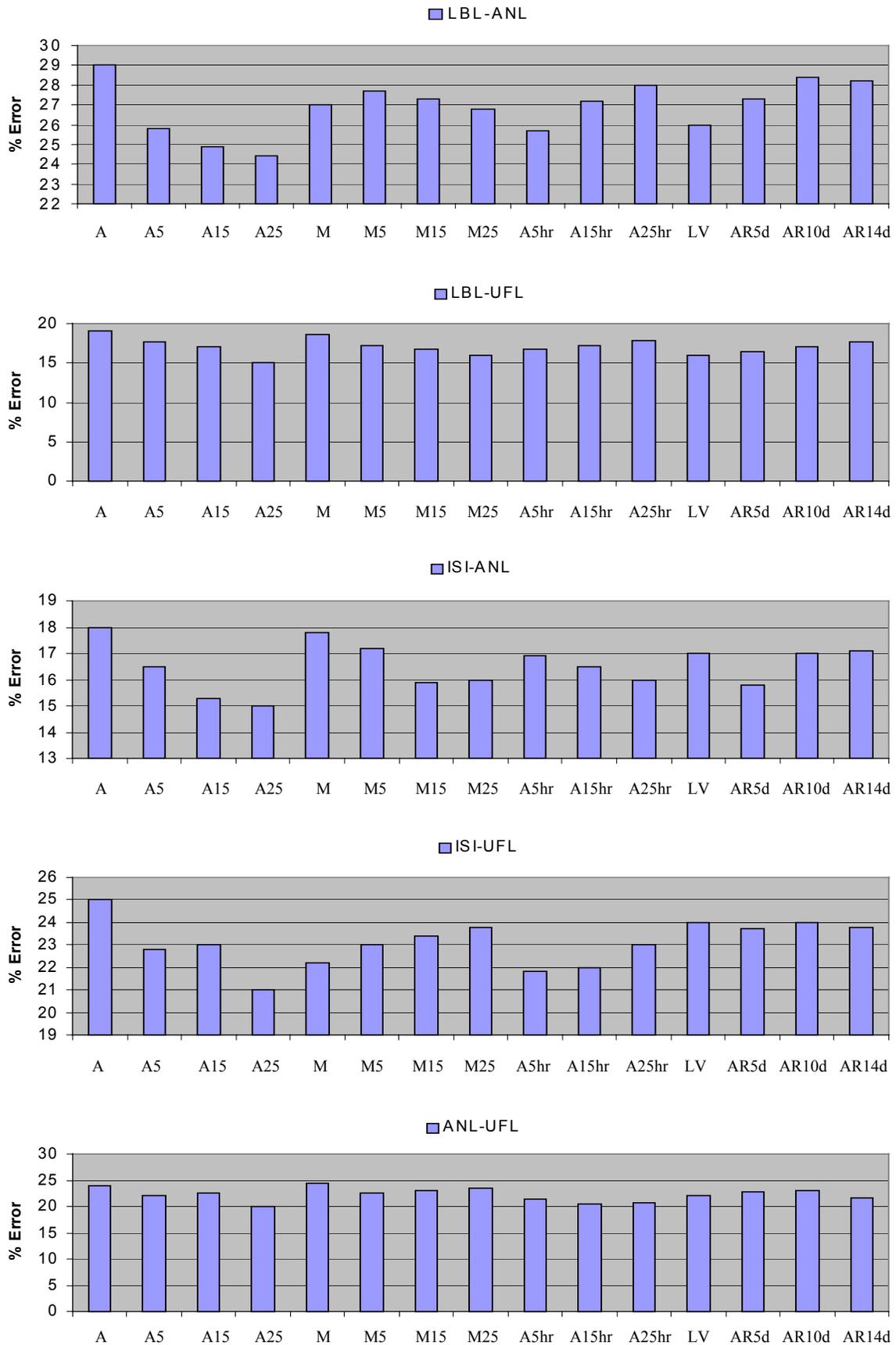

**Figure 6:** Univariate predictor performance for the testbed site pairs. Predictors include mean-based, median-based, and autoregressive models. The figure also shows context-insensitive variations of all the predictors.

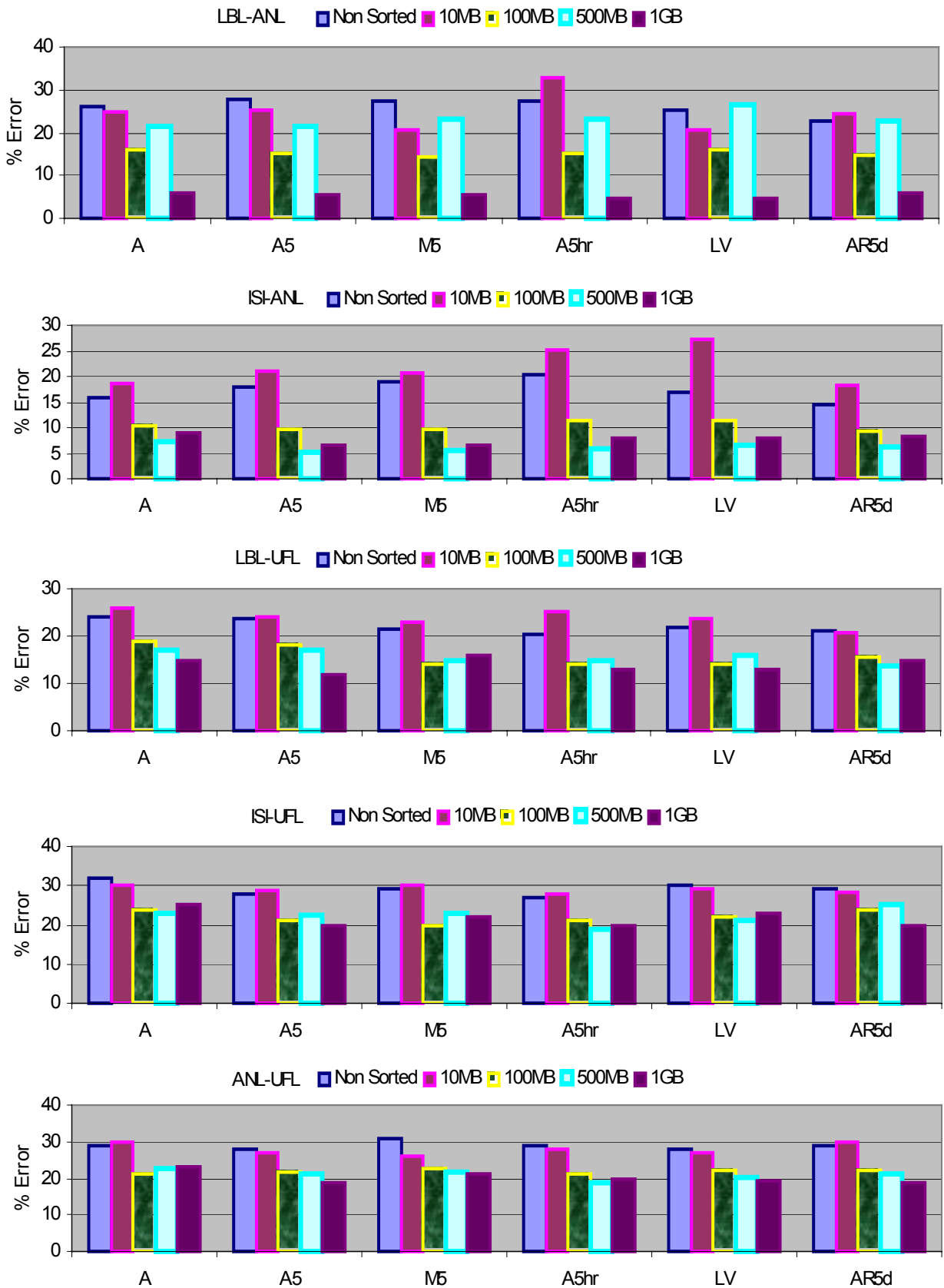

**Figure 7:** Impact of classification and the reduction in percent error rates for the testbed (context-sensitive filtering).

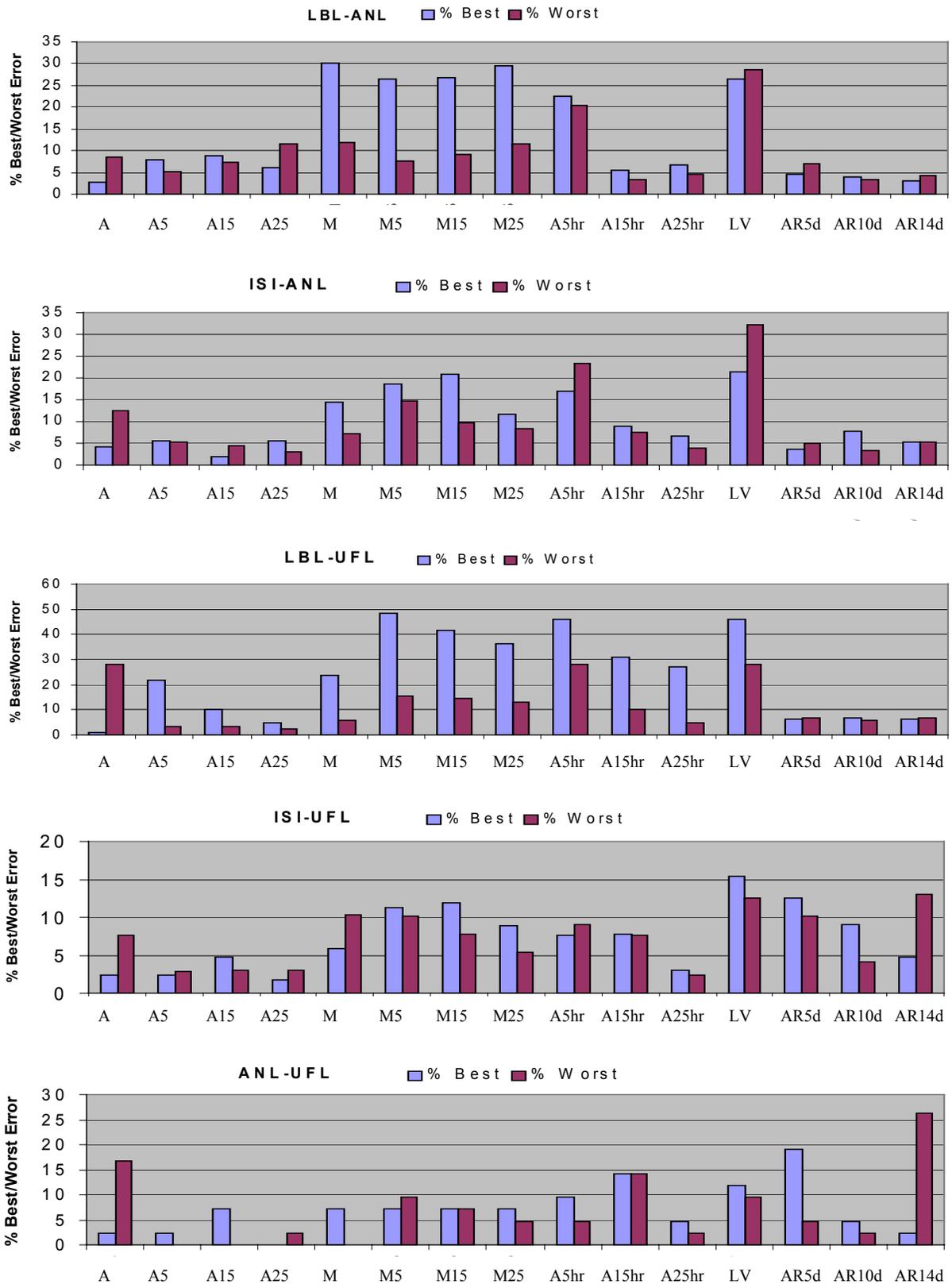

**Figure 8:** Relative performance of predictors as a percentage best/worst of all predictors for all site pairs.

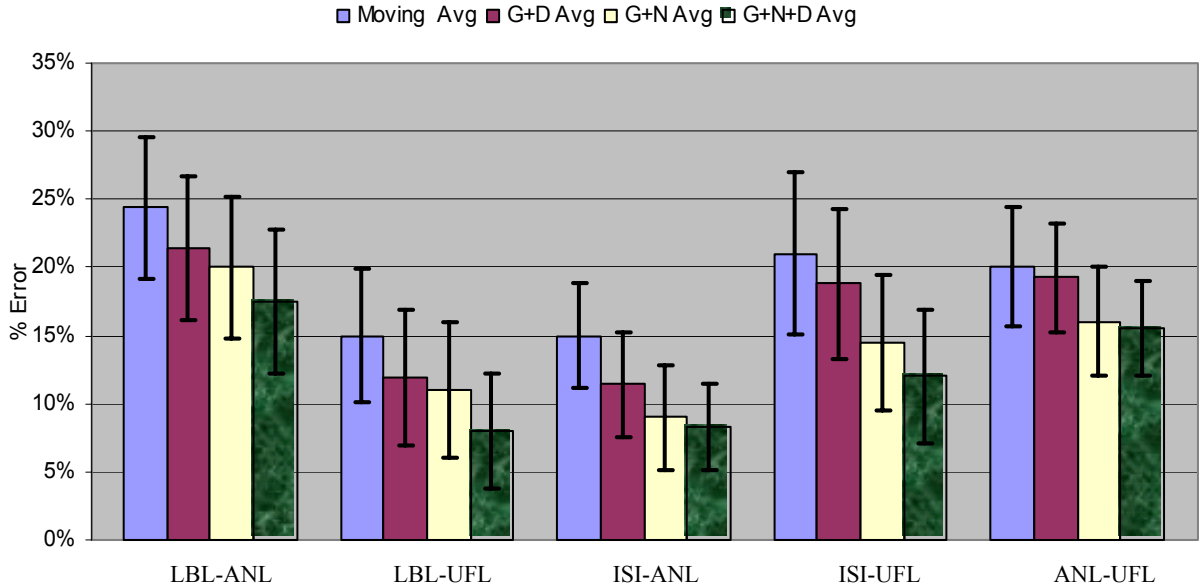

**(a) Comparison of normalized percent errors for the predictors with 95% confidence limits**

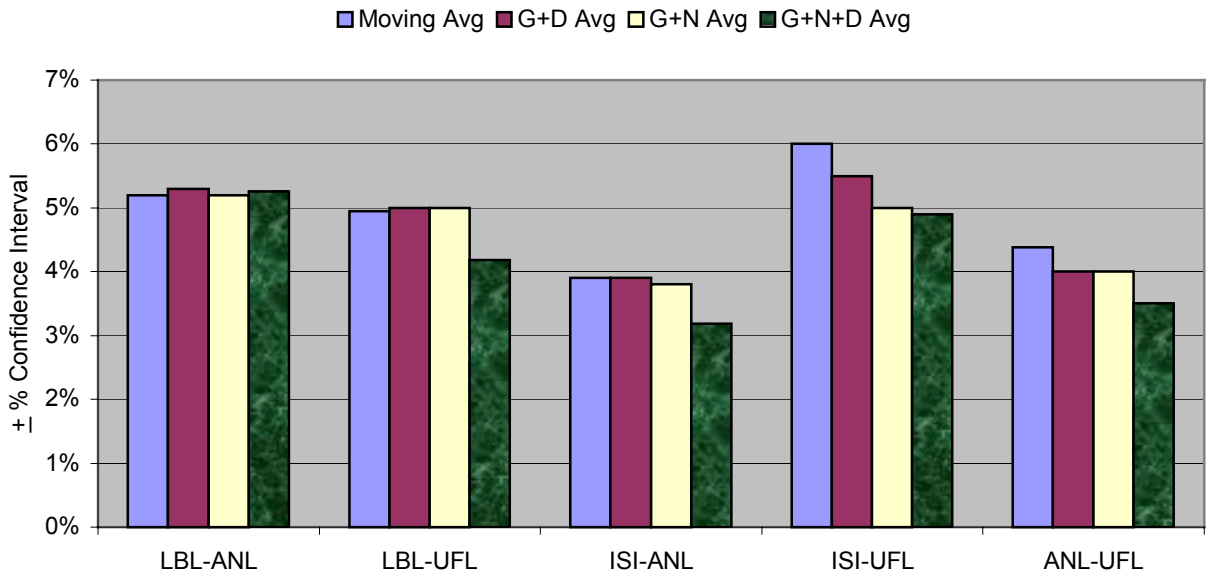

**(b) Comparison of intervals for the predictors**

**Figure 9:** (a) Normalized percent prediction error and 95% confidence limits for August 2001 dataset based on (1) prediction based on GridFTP in isolation (MovingAvg), (2) regression between GridFTP and disk I/O with Avg filling strategy (G+D Avg); (3) regression between GridFTP and NWS network data with Avg filling strategy (G+N Avg), and (4) regressing all three datasets (G+N+D Avg). Confidence Limits denote the upper and lower bounds of prediction error. For instance, the LBL-ANL pair had a prediction range of [17.3% $\pm$ 5.2%]. (b) Comparison of the percentage of variability among the predictors.

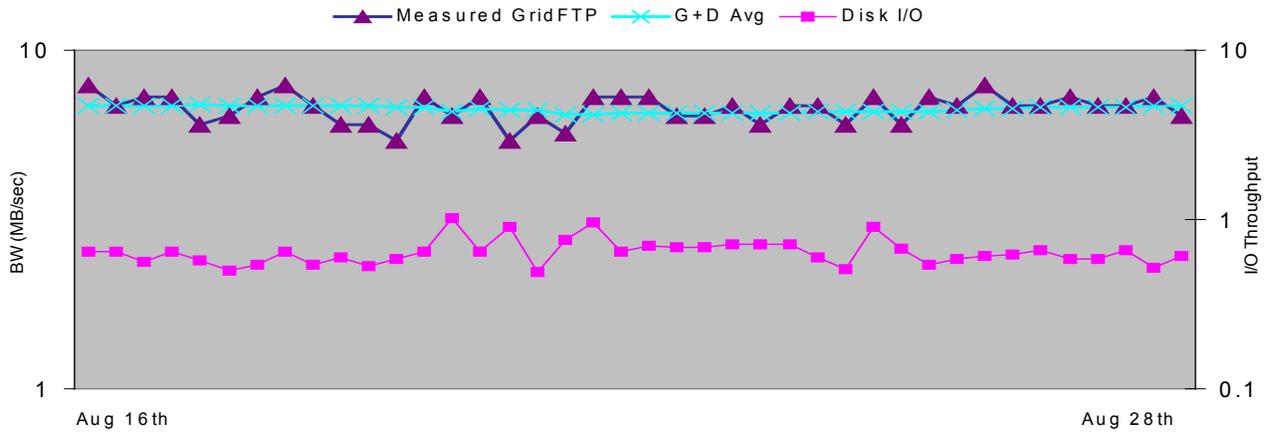

**(a) Regression between GridFTP and Disk I/O**

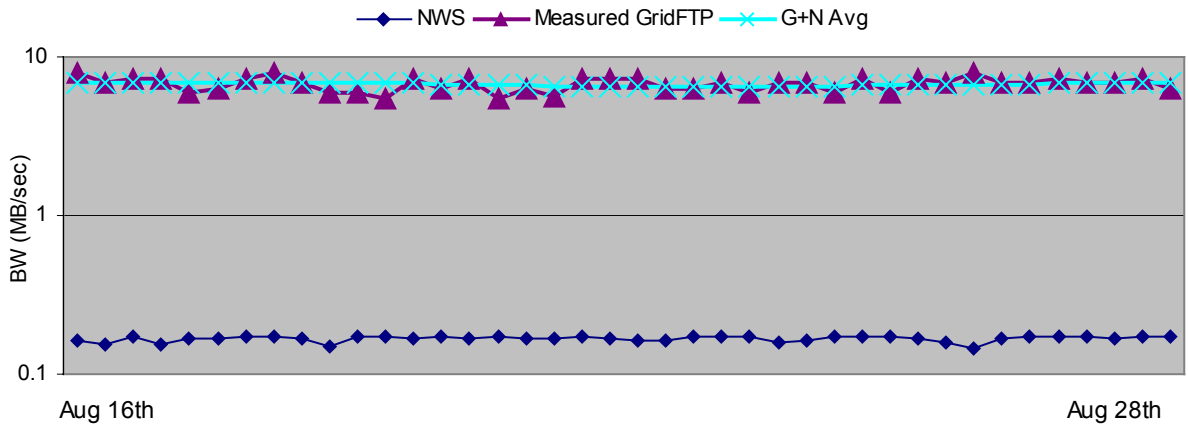

**(b) Regression between GridFTP and NWS**

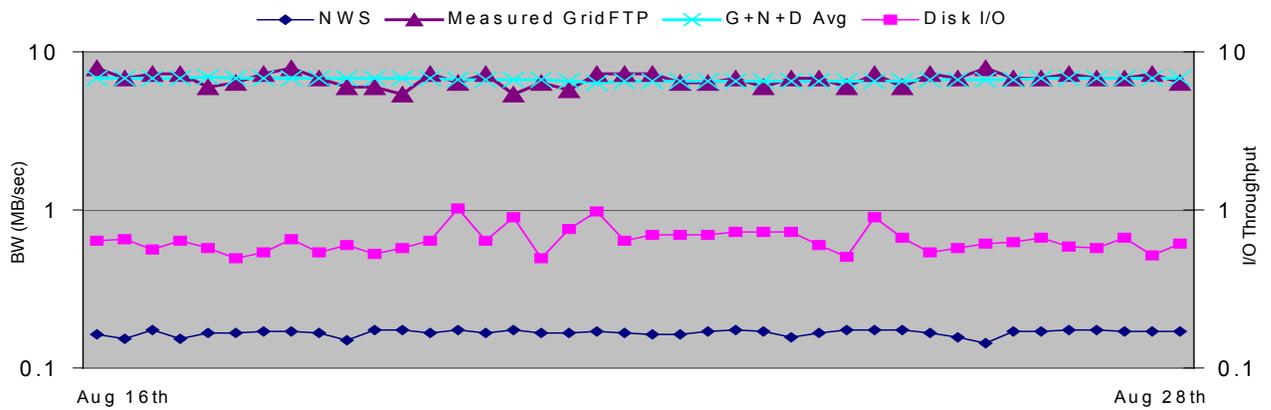

**(c) Regression between GridFTP, NWS and Disk I/O**

**Figure 10:** Predictors for 100 MB transfers between ISI and ANL for August 2001 dataset. In the graphs, GridFTP, G+D Avg, G+N+D Avg, and NWS are plotted on the primary *y*-axis; while Disk I/O is plotted on the secondary *y*-axis. I/O throughput denotes transfers per second.

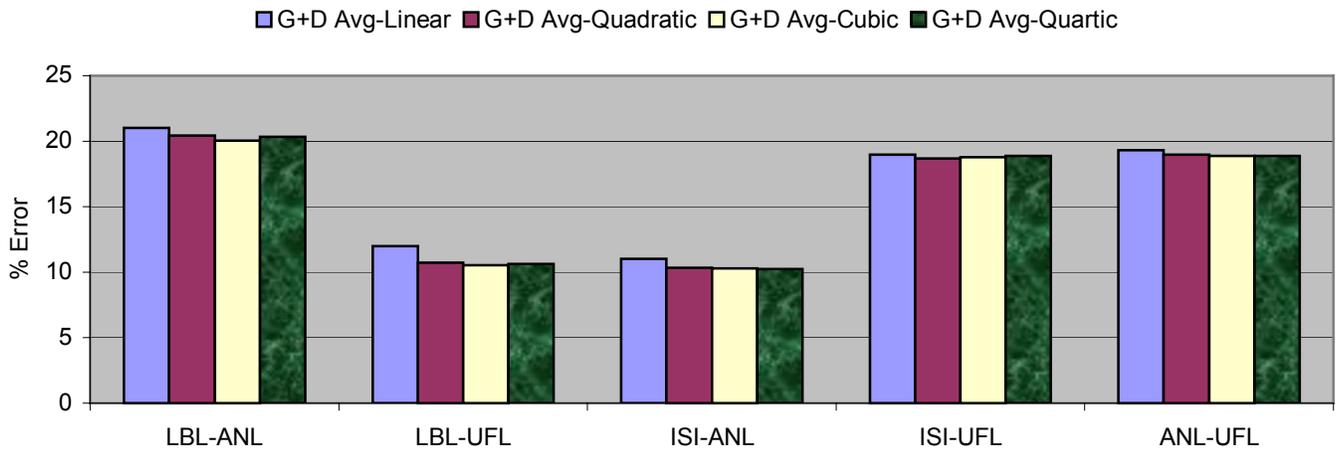

**Figure 11:** Error rates of polynomial regression models for the G+D Avg predictor for the various site pairs. Polynomials include linear, quadratic, cubic, and quartic models.